\documentclass[11pt]{article}
\usepackage{amsfonts,amssymb,amsmath,epsfig,graphicx}
\usepackage{mathrsfs}
\usepackage{theorem,calc} \usepackage{url} \usepackage{enumerate}
\usepackage{vmargin, cite}
\setmarginsrb{2cm}{2cm}{2cm}{2.5cm}{0cm}{0cm}{0cm}{0.5cm}

\usepackage{tikz} \usetikzlibrary{arrows}
\usetikzlibrary{positioning,arrows,chains,matrix,scopes,fit,decorations.markings,decorations}
\tikzstyle{block}=[draw opacity=0.7,line width=1.4cm]

\newtheorem{definition}{Definition}[section]

\newtheorem{proposition}[definition]{Proposition}

\newtheorem{lemma}[definition]{Lemma}
\newtheorem{rmk}[definition]{Remark}
\numberwithin{equation}{section}

\newcommand{\beq}{\begin{equation}} \newcommand{\eeq}{\end{equation}}
\newcommand{\bea}{\begin{eqnarray}} \newcommand{\eea}{\end{eqnarray}}

  \newcommand{\eeano}{\end{eqnarray*}}
\newcommand{\bma}{\begin{pmatrix}} \newcommand{\ema}{\end{pmatrix}}

\newcommand{\isp}{\psi^\ast}

\newcommand{\vps}{\varphi^\ast}
\newcommand{\vp}{\varphi}

\def \h#1{\widehat{#1}}
\def \wt#1{\widetilde{#1}}

\def \wth#1{\widehat{\widetilde{#1}}}

 \newcommand{\ZZ}{{\mathbb Z}}

\newcommand{\prf}{\underline{Proof:}\ } \newcommand{\finprf}{\null
  \hfill {\rule{5pt}{5pt}}}
\newcommand{\ie}{{\em ie.}~}

\newcommand{\bra}[1]{\langle #1 |}
\newcommand{\ket}[1]{| #1 \rangle}

\title{Eigenfunction equations of lattice KdV equations and \\ connections to ABS lattice equations with a  $\delta$ term\footnote{The authors are supported by NSFC (No.~11601312, 11671321, 11875040).}}
\date{\empty}
\author{   Cheng Zhang\footnote{Corresponding authors: ch.zhang.maths@gmail.com}\,,~~~~Haifei Zhang,~~~~Da-jun Zhang \\ \\
  \sc \small Department of Mathematics \\  \sc \small Shanghai University \\ \sc \small Shanghai, 200444, China
 }
\begin{document}
\maketitle
\begin{abstract}
  We develop lattice eigenfunction equations of lattice KdV equation, which are equations obeyed by the auxiliary functions, or eigenfunctions, of the Lax pair of the lattice KdV equation. 
  This leads to three-dimensionally consistent quad-equations that are closely related to lattice equations in the Adler-Bobenko-Suris (ABS) classification. In particular, we show how the H3($\delta$), Q1($\delta$) and Q3($\delta$) equations in the ABS list arise from the lattice eigenfunction equations by providing a natural interpretation of the $\delta$ term as interactions between the eigenfunctions. By construction, exact solution structures of these equations are obtained. The approach presented in this paper can be used as a systematic means to 
  search for integrable lattice equations.
\vspace{.2cm}

\end{abstract}

\section{Introduction}
Research in discrete integrable equations traces back in early developments of the modern theory of integrability to attempts of discretizing known soliton models. Systematic approaches includes, for instance,  Hirota's method by discretizing his bilinear equations \cite{HDE1, HDE2}; Date, Jimbo and Miwa's discretization of Sato's scheme \cite{DJM}; and the (discrete) direct linearization method, developed by Nijhoff, Quispel and Capel \cite{NQC}.  The common idea behind these methods is to add {\em discrete dynamics} to known integrable structures. 
This gave rise to lattice  versions of some well-known soliton equations.

In \cite{NQC}, a class of lattice Korteweg-de Vries (KdV) equations, as discrete analogs of continuous KdV-type equations, was introduced  (see also \cite{NAH, NC2, HJN1}). This includes the lattice potential KdV (lpKdV), lattice potential modified KdV (lpmKdV), lattice Schwarzian KdV (lSkdV), and Nijhoff-Quispel-Capel (NQC) equations.
They are  quad-equations, \ie lattice equations defined on quadrilaterals, and $\ZZ_2$-symmetric (symmetric by interchanging independent variables as well as the associated lattice parameters).
A unified framework was later developed, known as Cauchy matrix scheme \cite{NAH, HJN1},
containing all the above lattice KdV-type equations as well as their exact solutions.
It has been understood that lattice integrable models are rather {\em rare}, and can be regarded as ``master integrable models'',   since each contains information of the whole hierarchy of the associated continuous equation \cite{WC, NC2, HJN1}.

The notion of {\em three-dimensional consistency}, has then emerged as a defining criterion of integrability for quad-equations \cite{NW, Nijh2, BS3}. Namely, a quad-equation is said to be integrable, if it can be  consistently embedded onto a cube. The $3$D consistency implies that collective shifts, or {\em covariances}, of an integrable quad-equation, in a multi-dimensional lattice, are auto-B\"acklund transformations of itself.  This gives the very precise analog of infinitely many commuting flows in the continuous integrable field theory. 
An algorithm, up to few other assumptions, for solving the $3$D consistency was carried out by Adler, Bobenko and Suris, which led to a full classification of integrable quad-equations \cite{ABS1, ABS2}, known as the Adler-Bobenko-Suris (ABS) list. The list contains $9$ equations, most of them were known or related to known lattice KdV-type equations.
For example, the Q4 equation  \cite{AEqs, AS2}, as the ``top'' equation in the list, is the lattice version of the famous Krichever-Novikov (KN) equation \cite{KN1}; Q3, Q1, H3 and H1 are related to the NQC, lSKdV, lpmKdV and lpKdV equations respectively. However, there exist Q3($\delta$), Q1($\delta$) and H3($\delta$), which are new equations, and  apparently look like extensions with a constant term  $\delta$.  Many efforts have been made to understand the ABS list by studying their solution structures \cite{NAH, SQ3, HZ, NA-elliptic, NAcp,ZZ-SAPM-2013}. It turns out that  Q3($\delta$) is the ``second top'' equation in the list, as the remaining ones (apart from Q4) can be derived from it through certain degeneration processes \cite{ABS1, NAH, NA-elliptic, Bak}. The continuous limits of the ABS equations were also investigated in \cite{VerM}.

In the continuous realm, eigenfunction equations, \ie equation satisfied by the auxiliary function of a given Lax pair, are known to be natural extensions of some soliton models (see \cite{kono1} and references therein). For instance, the eigenfunction equations of KdV amounts to the ``vertical hierarchy'', introduced by Konopelcheko \cite{kono1}, which are equations derived from the Lax pair of KdV through successive Miura-type transformations. The hierarchy is degenerate, up to order two, and contains two main equations. By construction, their solution structures are expressed in terms of eigenfunctions, 
which can be written in compact forms thanks to the Darboux-Crum formulae \cite{Darboux, Crum, Darb2}.
A classical result in integrable systems is that the compatibility of B\"acklund transformations, also known as {\em nonlinear superposition formula},  of certain soliton equations yields integrable lattice equations \cite{Wah}. An alternative point of view consists in discretizing Lax pairs using Darboux transformations, {\em cf}.~\cite{LB1, LL, Shabat1}. This idea has been recently applied to equations in the ABS list, {\em cf}.~\cite{CaoZ, EKZ, ZhPeZh}, and associated Darboux-Crum formulae  were also provided \cite{ZhPeZh}. Note that the key ideas of the B\"acklund and/or Darboux transformation approaches as discretization processes, are  covariances of the underlying systems, which are in line with the notion of the $3$D consistency.

The aim of this paper is to develop eigenfunction equations of the lattice KdV equation as a means to  understand ABS lattice equations with a $\delta$ term. Based on the Lax formulation of the lattice KdV equation, we derive two lattice eigenfunction equations. They are $3$D-consistent,  closely related to H3 and Q3, and by construction inherit exact solution structures. Moreover, by some careful extension of the Lax pairs, we derive lattice eigenfunction equations with a constant $\delta$ term, where $\delta$ appears as an ``interacting'' term between different eigenfunctions. These equations give rise to the Q3($\delta$), Q1($\delta$), H3($\delta$) equations from the ABS list, and, as a result, exact solution structures of the latter equations can be constructed in terms of the eigenfunctions.
Our approach of constructing and solving Q3($\delta$), as well as Q1($\delta$) and H3($\delta$),  does not require presumed analytic properties of the solutions. 
The exact solution structures of Q3$(\delta)$ are also useful in investigating other equations, apart from Q4, in the ABS list through degeneration processes.
 The effectiveness of our approach suggests it can be used as a natural algorithm for  
searching for/extending lattice integrable models.

The paper is organized as follows. In Section $2$,  the Lax pair of the lattice KdV equation and associated Abel's formulae are provided. In Section $3$ and $4$, we present derivations of the lattice eigenfunction equations and their $\delta$-extensions. Their connections to known lattice equations, in particular, to Q3($\delta$), are also provided. 
Some concluding remarks are made in Section $5$.

\section{Lax pair of  lattice KdV  equation and discrete Abel's formulae}
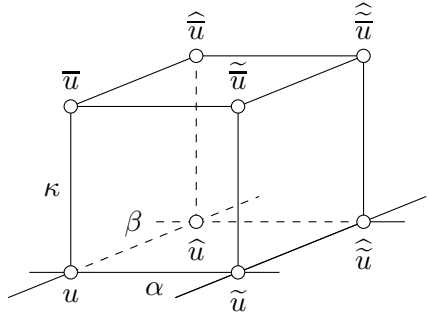
\begin{figure}[h]
	\centering
	\begin{tikzpicture}[scale=.55, decoration={markings,mark=at position 0.55 with {\arrow{latex}}}]
	\tikzstyle{nod1}= [circle, inner sep=0pt, fill=white, minimum size=5pt, draw]
	\tikzstyle{nod}= [circle, inner sep=0pt, fill=black, minimum size=5pt, draw]
	\def\lx{3}%
	\def\ly{1.22}%
	\def\lz{ (sqrt(\x*\x+\y*\y))}%
	\def\l{4}%
	\def\d{4}%
	\coordinate (u00) at (0,0);
	\coordinate (u10) at (\l,0);
	\coordinate (u01) at (\lx,\ly);
	\coordinate (u11) at (\l+\lx,\ly);
	\coordinate (v00) at (0,\d);
	\coordinate (v10) at (\l,\d);
	\coordinate (v01) at (\lx,\d+\ly);
	\coordinate (v11) at (\l+\lx,\d+\ly);
	\draw[-]  (u10)  
	-- (u11) ;
	\draw [dashed]  (u01)  -- (u11);
	\draw [dashed]  (u00)--node [above]{$\beta$}(u01) ;
	\draw[-] (v00) --  (v10) -- (v11) -- (v01) -- (v00);
	\draw[-] (u11) --  (v11);
	\draw[-] (u00) -- node [left]{$\kappa$}(v00);
	\draw[-] (u10) -- (v10);
	\draw[dashed] (u01) -- (v01);
	\coordinate (u011) at (1.5*\lx,1.5* \ly);
	\draw[dashed] (u01) -- (u011) ;
	\draw[dashed] (u01) -- (.66*\lx,\ly) ;
	\draw[-] (u00) -- (-0.5*\lx,-0.5* \ly);
	\draw[-] (u00) -- (-.33*\lx,0);
	\draw[-] (u10) -- (0.33*\lx+\l,0);
	\draw[-] (u10) -- (-.5*\lx+\l,-.5*\ly);
	\draw[-] (u11) -- (1.33*\lx+\l,\ly);
	\draw[-] (u11) -- (-.5*\lx+\l,-.5*\ly);
	\draw[-]  (u11)-- (1.5*\lx+\l,1.5* \ly);
	\draw[-] (u00) 
	-- node [below]{$\alpha$} (u10);
	\node[nod1] (v00) at (0,\d) [label=above: $\overline{u}$] {};
	\node[nod1] (v10) at (\l,\d) [label=above: $\widetilde{\overline{u}}$] {};
	\node[nod1] (v01) at (\lx,\d+\ly) [label=above: $\widehat{\overline{u}}$] {};
	\node[nod1] (v11) at (\l+\lx,\d+\ly) [label=above: $\widehat{\widetilde{\overline{u}}}$] {};
	\node[nod1] (u00) at (0,0) [label=below: $u$] {};
	\node[nod1] (u10) at (\l,0) [label=below: $\widetilde{u}$] {};
	\node[nod1] (u01) at  (\lx,\ly) [label=below: $\widehat{u}$] {};
	\node[nod1] (u11) at (\l+\lx,\ly) [label=below: $\widehat{\widetilde{u}}$] {};
	\end{tikzpicture}
	\caption{$3$D consistency: here $\overline{u}$ denotes shift of $u$ in the third direction.} \label{fig:31}
\end{figure}

We first collect some basic notions and useful notations.  Let $u$ be a discrete field depending on $n,m \in \ZZ$. Shifts in $n$ and $m$  are denoted by ~$\widetilde{~}$~  and ~$\widehat{~}$~  as
\begin{equation}
  \label{eq:2}
  \widetilde{u}(n,m) = u(n+1,m)\,,\quad  \widehat{u}(n,m) = u(n,m+1)\,,\quad   \widehat{\widetilde{u}}(n,m) = u(n+1,m+1)\,, \quad \dots
\end{equation}
As an illustration, the lpKdV equation reads
\begin{equation} \label{eq:H1}
  (  \widetilde{u} -\widehat{u})(\widehat{\widetilde{u}}-u) =\alpha^2-\beta^2\, ,
\end{equation}
where  $\alpha$ and  $\beta$ are lattice parameters associated to the ~$\widetilde{~}$ and  ~$\widehat{~}$~ directions respectively. Thus, $u$ also depends on the lattice parameters. The $3$D consistency  is depicted in Fig.~\ref{fig:31},  where $u$ depends on three, or more, discrete variables, and also on the associated lattice parameters.

Now, consider the following system of linear difference equations
\begin{subequations}
  \label{eq:dschreq}  \begin{align}
    \label{eq:dschreq1}
    L\,\varphi & = \widetilde{\widetilde{\varphi}} +h\widetilde{\varphi}+\alpha^2\,\varphi = p^2
                 \varphi\,,       \\
    \label{eq:dschreq2}
    T\,\varphi & =  \widehat{\varphi} =\widetilde{\varphi}-g\varphi \,,
  \end{align}
  \end{subequations}
where  $h, g$ are discrete fields of $n, m\in \ZZ$, and $\alpha$ is the lattice parameter associated to the  ~$\widetilde{}$~  direction. The discrete spectral problem \eqref{eq:dschreq1} can be obtained through {\em exact discretization} \cite{Shabat1} of the continuous Schr\"odinger spectral problem, \ie $(D^2+u)\varphi = p^2 \varphi$, and  \eqref{eq:dschreq2} is the discrete iso-spectral deformation of $L$
. They form the Lax pair for the lattice KdV equation \cite{HJN1, ZhPeZh}.
\begin{proposition}\label{prop:1}
The compatibility condition for  the system \eqref{eq:dschreq}, \ie $\widehat{\widetilde{\widetilde{\varphi}}}=\widetilde{\widetilde{\widehat{\varphi}}}$, 
yields, up to one step of integration in the  ~$\widetilde{}$~  direction, the following system,
\begin{equation}
    \label{eq:10sys}
    \widehat{h}-\widetilde{h} =\widetilde{\widetilde{g}}-g \,, \quad (h+\widetilde{g})g =\beta^2 -\alpha^2\,,
\end{equation}
  where $\beta^2 -\alpha^2$ is the constant of integration. Moreover, $\varphi$ satisfies
\begin{equation}\label{eq:hvarph}
 \widehat{\widehat{\varphi}} +\eta\widehat{\varphi}+\beta^2\varphi = p^2 \varphi\,,\quad \eta=  h+\widetilde{g}+\widehat{g}\,.
\end{equation}
\end{proposition}

The proof can be verified by straightforward computations. The constant of integration will be fixed to be $\beta^2 -\alpha^2$ in the rest of paper, so that  $\beta$ plays the role of lattice parameter associated to the ~$\widehat{}$~ direction.
By eliminating $h$, the system \eqref{eq:10sys} involving $h, g$ defines the lattice KdV equation \cite{HDE1, NC2, HJN1}
\begin{equation}
  \widetilde{f}-\widehat{f}+ \widehat{\widetilde{g}} -g = 0\,, \quad fg=\beta^2-\alpha^2\,,
\end{equation}
with $f=h+\widetilde{g}$. Introduce a ``potential'' variable $u$ obeying
\begin{equation}
  \label{eq:1}
  g =  \widehat{u}-\widetilde{u}\,,\quad h=\widetilde{\widetilde{u}}-u\,,
\end{equation}
then, the system \eqref{eq:10sys} is reduced to the lpKdV equation \eqref{eq:H1}.

Under the relation \eqref{eq:1}, the two  spectral problems  \eqref{eq:dschreq1} and \eqref{eq:hvarph}, connected by \eqref{eq:dschreq2}, are in the same forms by interchanging (~$\widetilde{}$~, $\alpha$) and (~$\widehat{}$~, $\beta$). This property provides an apparent evidence that the lpKdV equation \eqref{eq:H1} is $3$D-consistent: let $\overline{\varphi} = \widetilde{\varphi}-(\overline{u}-\widetilde{u})\varphi$ be the deformation of $L$ in the third direction, namely, the ~$\bar{~}$~ direction, then  $\overline{\varphi} = \widehat{\varphi}-(\overline{u}-\widehat{u})\varphi$, which is also the ``right'' deformation of \eqref{eq:hvarph}. Therefore, the three Lax pairs among any two of the ~$\widetilde{}$~,    ~$\widehat{}$~,  ~$\bar{~}$~  directions are compatible, and yield the same equations. Covariances of the Lax pairs create a $3$D object, \ie a cube, whose faces are governed by the same quad-equations.

\begin{rmk}
For simplicity, we only consider the autonomous case, \ie  the lattice parameters $\alpha$ and $\beta$ are set to be constant. However, Prop.~\ref{prop:1} also holds in the non-autonomous case by letting dependence of  $\alpha$ (resp.~$\beta$) on $n $ (resp.~$m$). Then, the lattice KdV and lpKdV equations become non-autonomous.
\end{rmk}

A direct consequence of Prop.~\ref{prop:1} for the auxiliary field  $\varphi$ is  the following.
\begin{lemma}[Discrete Abel's formulae]\label{lem1}Let  $\varphi_1, \varphi_2$ be two linearly independent solutions of the Lax pair \eqref{eq:dschreq}, then
  \begin{equation}
    \label{eq:9}
    \varphi_1\widetilde{\varphi}_2 -    \varphi_2\widetilde{\varphi}_1 =    \varphi_1\widehat{\varphi}_2 -    \varphi_2\widehat{\varphi}_1 = \rho \, s^{2n}t^{2m}\,,
  \end{equation}
  where $\rho$ is a constant of integration, and  $s,t $ satisfy
  \begin{equation}
  \label{eq:5para}
   s^2 = \alpha^2 -p^2\,,\quad t^2 = \beta^2 - p^2\,.
\end{equation}
\end{lemma}
\prf The quantities  $\varphi_1\widetilde{\varphi}_2 -    \varphi_2\widetilde{\varphi}_1$ and $ \varphi_1\widehat{\varphi}_2 -    \varphi_2\widehat{\varphi}_1$ are the Casorati determinant of $\varphi_1, \varphi_2$ along the $~\widetilde{~}~$  and  $~\widehat{~}~$ directions respectively. We employ the following notations
\begin{equation}
  \label{eq:12}
  | \varphi_1,  \varphi_2|_{\widetilde{~}} := \varphi_1\widetilde{\varphi}_2 -    \varphi_2\widetilde{\varphi}_1 \,,\quad  | \varphi_1,  \varphi_2|_{\widehat{~}} :=   \varphi_1\widehat{\varphi}_2 -    \varphi_2\widehat{\varphi}_1\,.
\end{equation}
Thanks to  \eqref{eq:dschreq2}, these two quantities are equal. It follows from \eqref{eq:dschreq1} and \eqref{eq:hvarph} that 

\begin{equation}
  \label{eq:15t}
  \widetilde{ | \varphi_1,  \varphi_2|_{\widetilde{~}} } =  (\alpha^2 - p^2) | \varphi_1,  \varphi_2|_{\widetilde{~}} \,,\quad \widehat{ | \varphi_1,  \varphi_2|_{\widehat{~}} }=   (\beta^2 - p^2) | \varphi_1,  \varphi_2|_{\widehat{~}}\,.
\end{equation}
Solving these equations completes the proof. \finprf
\medskip

Lemma~\ref{lem1} is a discrete analog of the Abel's formula for linear differential equations. In the rest of the paper, we will use $\{\varphi_1, \varphi_2\}$ to denote a solution basis of the solution spaces of the Lax pair \eqref{eq:dschreq}. For an nonzero $p$, a natural choice could be
\begin{equation}
  \label{eq:sb1}
\varphi_1=\varphi(n,m;p)\,,\quad \varphi_2=\varphi(n,m;-p)  \,.
\end{equation}
\begin{rmk}\label{rmk:DTs}
Note that classes of exact solutions for the Lax pair \eqref{eq:dschreq} can be constructed following an algebraic approach, known as Darboux-Crum transformation \cite{ZhPeZh}. The function $\varphi, h$ admit compact expressions as ratios of determinants under the actions of Darboux transformations. In Appendix~\ref{app:1},  
explicit formulae for $\varphi$ and $h$ for soliton solutions 
and elliptic soliton solutions are provided.
\end{rmk}


\section{Lattice eigenfunction KdV equations}
This section aims to construct eigenfunction equations of the Lax pair \eqref{eq:dschreq}. We obtain the following two quad-equations
\begin{equation}
\label{H3eq}
(\alpha^2-p^2)\varphi{\widehat{\varphi}} -(\beta^2-p^2)\varphi{\widetilde{\varphi}} - \widetilde{\varphi}\widehat{\widetilde{\varphi}}+ \widehat{\varphi}\widehat{\widetilde{\varphi}}=0\,,
\end{equation}
and
\begin{equation}
\label{H3eqdd}
(\alpha^2 -p^2)\psi{\widehat{\psi}} -  (\beta^2 - p^2)\psi{\widetilde{\psi}}- \widetilde{\psi}\widehat{\widetilde{\psi}}+ \widehat{\psi}\widehat{\widetilde{\psi}}=(-1)^{n+m}\delta (\alpha^2 -\beta^2 )(\alpha^2-p^2)^{n}(\beta^2-p^2)^{m}\,,
\end{equation}
where $\delta$ is a constant. For obvious reasons, we call \eqref{H3eq}  {\em lattice eigenfunction KdV}  (leKdV) equation  and \eqref{H3eqdd}  leKdV($\delta$) equation.  Apparently, leKdV is a special case of leKdV($\delta$) with $\delta=0$. Both equations are $3$D-consistent, but only $\ZZ_2$-symmetric, \ie symmetric in interchanging   $(~\widetilde{~}~, \alpha)$ and $(  ~\widehat{~}~,\beta)$. 
They are closely related to H3 and H3($\delta$).
\subsection{Derivations}
The leKdV equation \eqref{H3eq} can be easily obtained by eliminating $h,g$ from the  \eqref{eq:dschreq} and  \eqref{eq:10sys}. Note that there is a ``free'' parameter $p$ appearing in the equation
.  Although it is an nonlinear equation, any linear combination of $\varphi$'s, as solutions to  \eqref{eq:dschreq},  still solves \eqref{H3eq}.  The lattice KdV equation and leKdV are connected through Miura-type transformations
\begin{equation}
  \label{eq:4miura}
  g = \frac{\widetilde{\varphi} -\widehat{\varphi} }{\varphi}\,,\quad   h+\widetilde{g} = \frac{(p^2-\alpha^2)\varphi -\widehat{\widetilde{\varphi}} }{\widetilde{\varphi}}\,.
\end{equation}

Here, we provide an  alternative approach to constructing  leKdV by using the discrete Abel's formulae stated in Lemma~\ref{lem1}.
\begin{proposition}\label{prop:31}
The discrete Abel's formulae \eqref{eq:9} define auto-B\"acklund transformations between $\varphi_1$ and $\varphi_2$ as solutions to the leKdV equation \eqref{H3eq}.
\end{proposition}
\prf The formulae \eqref{eq:9} amount to
\begin{subequations}
  \label{eq:sysm10}
  \begin{align}
    \varphi_1\widetilde{\varphi}_2 -    \varphi_2\widetilde{\varphi}_1 =  \rho\, s^{2n}t^{2m}\,, \quad &    \widehat{\varphi}_1\widehat{\widetilde{\varphi}}_2 -  \widehat{  \varphi}_2\widehat{\widetilde{\varphi}}_1 =  \rho\, s^{2n}t^{2m+2}\,,\\
\varphi_1\widehat{\varphi}_2 -    \varphi_2\widehat{\varphi}_1 = \rho\, s^{2n}t^{2m}        \,, \quad & \widetilde{  \varphi}_1\widehat{\widetilde{\varphi}}_2 - \widetilde{   \varphi}_2\widehat{\widetilde{\varphi}}_1 = \rho\, s^{2n+2}t^{2m}\,.
  \end{align}
  \end{subequations}
  By eliminating, for instance, $\widetilde{\varphi}_2, \widehat{\varphi}_2$, one has
  \begin{equation}
(\varphi_1   \widehat{\widetilde{\varphi}}_2    -\varphi_2 \widehat{\widetilde{\varphi}}_1)  \widehat{\varphi}_1 =  (  t^2 \varphi_1+ \widehat{\widetilde{\varphi}}_1)  \rho s^{2n}t^{2m} \,,\quad (\varphi_1   \widehat{\widetilde{\varphi}}_2    -\varphi_2 \widehat{\widetilde{\varphi}}_1)  \widetilde{\varphi}_1 =  (  s^2 \varphi_1+ \widehat{\widetilde{\varphi}}_1)  \rho s^{2n}t^{2m}\,,   \end{equation}
which turns out to be the leKdV equation for $\varphi_1$ by eliminating the term 
$(\varphi_1   \widehat{\widetilde{\varphi}}_2    -\varphi_2 \widehat{\widetilde{\varphi}}_1) $. 
Similarly, $\varphi_2$ also solves leKdV.
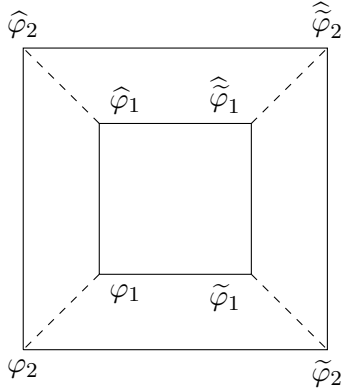
\begin{figure}[ht]
  \centering
  \begin{tikzpicture}[scale=.5, decoration={markings,mark=at position 0.55 with {\arrow{latex}}}]
 \def\l{4}%
    \def\d{2}%
       \coordinate (u00) at (0,0);
    \coordinate (u10) at (\l,0);
    \coordinate (u01) at (0,\l);
    \coordinate (u11) at (\l,\l);
    \coordinate (v00) at (-\d,-\d);
    \coordinate (v10) at (\l+\d,-\d);
    \coordinate (v01) at (-\d,\l+\d);
    \coordinate (v11) at (\l+\d,\d+\l);
    \draw[-] (u00) node[below right] {$\varphi_1$} -- (u10) node[below left] {$\widetilde{\varphi}_1$} -- (u11)  node[above left] {$\widehat{\widetilde{\varphi}}_1$} -- (u01) node[above right] {$\widehat{\varphi}_1$} -- (u00);
    \draw[-] (v00) node[below] {$\varphi_2$} --  (v10) node[below] {$\widetilde{\varphi}_2$} -- (v11)  node[above] {$\widehat{\widetilde{\varphi}}_2$} -- (v01) node[above] {$\widehat{\varphi}_2$} -- (v00);
        \draw[dashed] (u00) -- (v00);
        \draw[dashed](u10) -- (v10);
        \draw[dashed](u01) -- (v01);
        \draw[dashed](u11) -- (v11);
  \end{tikzpicture}
  \caption{Auto-B\"acklund transformation as consistency between $\varphi_1$ and $\varphi_2$. } \label{fig:back}
\end{figure}
\finprf

\medskip
In other words,  $\varphi_1$, $\varphi_2$ and their auto-B\"acklund transformations \eqref{eq:sysm10} form a $3$D-consistent system as shown in Fig.~\ref{fig:back}. To proceed to the $\delta$-extension of the leKdV equation, one needs to introduce an  ``adjoint'' Lax pair
\begin{equation}
  \label{eq:dschreqa}
     \widetilde{\widetilde{\vps}} -h\widetilde{\vps}+\alpha^2\,\varphi^\ast = p^2
                 \vps\,,       \quad      \widehat{\vps} =\widetilde{\vps}+g\vps\,.
  \end{equation}
  Here,  $h,g$ are the same functions as appeared in \eqref{eq:dschreq}. Similar results stated in Prop.~\ref{prop:1} and Lemma~\ref{lem1} also hold for the adjoint Lax pair. One can view $\vps$  as a Darboux transformation of $\vp$ as
  \begin{equation}
  \vps =  (-1)^{n+m}\varphi  \,.
  \end{equation}
\begin{lemma}\label{lem2}Let $\varphi$ and $ \varphi^*$ be solutions of the Lax pairs \eqref{eq:dschreq} and \eqref{eq:dschreqa} respectively, then
  \begin{equation}
    \label{eq:BK2}
    \varphi \widetilde{\varphi}^* +    \varphi^*\widetilde{\varphi} =    \varphi \widehat{\varphi}^* +    \varphi^*\widehat{\varphi} =  (-1)^{n+m}\rho\, s^{2n}t^{2m}\,,
  \end{equation}
  where $\rho$ is a constant of integration, and  $s, t $ are defined in (2.9).
 If $\vps \propto (-1)^{n+m}\vp$, then $\rho=0$.
\end{lemma}
The proof is straightforward. Similar to Prop.~\ref{prop:31}, the relations \eqref{eq:BK2}  define auto-B\"acklund transformations between  $\vp$ and $\vps$ both solving the leKdV equation. However, their linear combinations are solutions to the leKdV($\delta$) equation.
\begin{proposition}\label{propD3}
  Let $\psi = C_1 \varphi +C_2 \vps$, $\isp = C_1 \varphi -C_2 \vps$, where  $C_1, C_2$ are nonzero constants, and  $\varphi$ and $\vps$ solve respectively the Lax pairs \eqref{eq:dschreq} and \eqref{eq:dschreqa}. Then, $\psi$ and $\isp$ solve the leKdV($\delta$) equation \eqref{H3eqdd}. In particular,  $\delta =0 $ if $\vps \propto (-1)^{n+m}\vp$; otherwise, $\delta =2\rho  C_1 C_2$, where $\rho$ in the constant of integration appeared in \eqref{eq:BK2}.
\end{proposition}
\prf
It follows from Lemma~\ref{lem2} that
\begin{equation}\label{3.10}
  \psi \widetilde{\psi}-\isp\widetilde{\isp} =
  \psi \widehat{\psi}-\isp\widehat{\isp} =2C_1C_2(  \varphi \widetilde{\varphi}^* +    \varphi^*\widetilde{\varphi} ) = 2\rho\, C_1C_2(-1)^{n+m}s^{2n}t^{2m}\,,
\end{equation}
where  $s, t $ are defined in (2.9).
The constant of integration  $\rho$ vanishes, if $\vps \propto (-1)^{n+m}\vp$. Let  $\delta = 2 \rho\, C_1C_2 $, then the above relations after shifts yield
\begin{equation}
\widehat{\psi} \widetilde{\widehat{\psi}}-\widehat{\isp}\widetilde{\widehat{\isp}} =  \delta\,(-1)^{n+m+1}s^{2n}t^{2m+2}\,,\quad    \widetilde{\psi} \widehat{\widetilde{\psi}}-\widetilde{\isp}\widehat{\widetilde{\isp}} =\delta\, (-1)^{n+m+1}s^{2n+2}t^{2m}\,.
\end{equation}
By eliminating, for instance,  $\widetilde{\isp}$,   $\widehat{\isp}$, one gets
\begin{equation}
  \label{eq:d}
  \frac{\widehat{\widetilde{\isp}}}{\isp}=\frac{\widehat{\psi}\widehat{\widetilde{\psi}} +  \delta(-1)^{n+m}s^{2n}t^{2m+2}  }{\psi\widehat{\psi} - \delta(-1)^{n+m}s^{2n}t^{2m}}=\frac{\widetilde{\psi}\widehat{\widetilde{\psi}} +  \delta(-1)^{n+m}s^{2n+2}t^{2m}  }{\psi\widetilde{\psi} - \delta(-1)^{n+m}s^{2n}t^{2m}}\,.
\end{equation}
The second equality coincides with leKdV($\delta$)  for $\psi$. Similarly,    $\isp$  also  solves  leKdV($-\delta$).
\finprf
\medskip
\begin{rmk}
The function $\psi$ and $\isp$ in Prop.~\ref{propD3} satisfy a coupled linear system
\begin{equation}
  \widetilde{\widetilde{\Psi}}+ \bma  0 &  h \\ h & 0\ema \widetilde{\Psi} +\alpha^2\Psi
  =p^2\Psi\,, \quad  \widehat{\Psi} = \widetilde{\Psi} +\bma  0 &  g \\ g & 0\ema \Psi\,, \quad \Psi:= \bma \psi \\ \isp \ema\,. \end{equation}
The leKdV($\delta$) equation can be seen as the eigenfunction equation of this coupled system.
\end{rmk}

From the above construction, we could conclude that the leKdV and leKdV($\delta$) equations share the same solution basis, and the term $\delta$ is a result of ``interactions'' between the eigenfunction $\vp$ and its adjoint $\vps$ (comparing \eqref{eq:9} and \eqref{eq:BK2}).

Without loss of generality,  the most general form of $\psi$ can be written as
\begin{equation}
\label{eq:ABCD}   \psi = A \vp_1 +B \vp_2+C \vps_1+D \vps_2\,, \quad \vps_1 = (-1)^{n+m}\vp_1\,,\quad \vps_2 = (-1)^{n+m}\vp_2\,,
 \end{equation}
 where $\vp_1, \vp_2$ (resp.~ $\vps_1, \vps_2$ ) form a basis of the solution spaces of \eqref{eq:dschreq} (resp.~\eqref{eq:dschreqa}). Then $\psi$ satisfies the leKdV($\delta$) equation with $\delta \propto ( A D -B C)$.

 \subsection{Connections to H3($\delta$) and B\"acklund transformations}
Now we connect the leKdV($\delta$) equation to $\delta$-extensions of some well-known $3$D-consistent equations. Note that the parametrization \eqref{eq:5para} causes multivalueness of  $s,t$, which are  functions of $p$. In practice, we can fix the branch of the square root with  the sign of $p$ as
\begin{equation}\label{sign1}
  \text{sign} \,s(p) = \text{sign}\,  p\,, \quad   \text{sign} \,t(p) = \text{sign}\,  p\,.
\end{equation}

\begin{itemize}
\item Connection to lpmKdV($\delta$) equation: let $\psi$ satisfy the  leKdV($\delta$) equation  \eqref{H3eqdd}, and
\begin{equation}
  \psi  = s^{n}t^{m} v\,, \quad v:=v(n,m;s,t)\,,
\end{equation}
then $v$ satisfies a $\delta$-extension of the lpmKdV equation
\begin{equation}
  \label{eq:lpmkdvd1}
 s (v\widehat{v}  -\widetilde{v}\widehat{\widetilde{v}})-t(v\widetilde{v}  -\widehat{v}\widehat{\widetilde{v}}) = (-1)^{n+m}\delta\left(\frac{s}{t}-\frac{t}{s}\right)\,,
\end{equation}
which is 3$D$-consistent and $\ZZ_2$-symmetric. When $\delta=0$, this equation is reduced to the lpmKdV equation \cite{NC2}.
\item Connection to H3($\delta$): let $\psi$ satisfy the  leKdV($\delta$) equation \eqref{H3eqdd}, and
  \begin{equation}\label{eq:psii}
  \psi  =  (i\,s)^{n}(i\,t)^{m} w\,, \quad w:=w(n,m;s,t)\,,
\end{equation}
where $i$ denotes  the  imaginary unit, then $w$ satisfies
\begin{equation}\label{mh3ddd}
 t(w\widetilde{w} +\widehat{w}\widehat{\widetilde{w}})-s (w\widehat{w} +\widetilde{w}\widehat{\widetilde{w}}) = i\,\delta\left(\frac{s}{t}-\frac{t}{s}\right)\,,
\end{equation}
which is the H3($\delta$) equation from the ABS list \cite{ABS1}.
\end{itemize}
 \begin{rmk}
It follows from \eqref{eq:ABCD} that exact solutions of the H3($\delta$) can be expressed  by
\begin{equation}\label{eq:wis}
   w = (i\,s)^{-n}(i\,t)^{-m} \left((A+(-1)^{n+m}C )\vp_1 +(B +(-1)^{n+m}D)\vp_2\right)\,,
 \end{equation}
 where $\delta \propto AD-BC$. Similar solution structures for  H3($\delta$)  were found in \cite{NAH} for soliton solutions and in \cite{NA-elliptic} for elliptic soliton solutions, as degenerate cases of the exact solutions of  Q3($\delta$).  Here,  by construction,  exact solution structures of H3($\delta$) are in the above forms, and do not rely on particular analytic structures of the solution basis.
 \end{rmk}

 The Abel's formulae presented in the previous Section enable us to establish some nontrivial auto-B\"acklund transformations of the lattice equations \eqref{eq:lpmkdvd1} and \eqref{eq:wis}. Take  lpmKdV (equation \eqref{eq:lpmkdvd1} with $\delta=0$) as an example. Using
\begin{equation}
\varphi_1=s^nt^m V,~~ \varphi_2=s^nt^m U,
\end{equation}
where $\vp_1$, $\vp_2$ are two linearly independent solutions of the Lax pair \eqref{eq:dschreq}, then \eqref{eq:9} is reduced to
\begin{equation}\label{BT-uU}
 V \wt U-U \wt V=\rho/s,\quad  V\h U-U \h V=\rho/t.
\end{equation}
The compatibility conditions $\wt{\h V}=\h{\wt V}$ and $\wt{\h U}=\h{\wt U}$  yield two separated lpmKdV equations for $V$ and $U$.

The above structures form a so-called {\em consistent triplet}, which can be seen as an extended version of the $3$D consistency. Having a relation $H$ connecting $V, \wt{V}$ and $U, \wt{U}$ (also  $V, \h{V}$ and $U, \h{U}$)
\begin{equation}
 H(V,\wt{V},U, \wt{U},p)=0,\quad  H(V,\h{V},U,\h{U},q)=0,
\end{equation}
if the compatibility conditions $\wt{\h V}=\h{\wt V}$ and $\wt{\h U}=\h{\wt U}$ amount to two separated equations for $V$ and $U$
\begin{equation}
F(V,\wt V,\h V, \wth V; p,q)=0,\quad G(U,\wt U,\h U, \wth U; p,q)=0,
\label{u-eq}
\end{equation}
then $H, F, G$ form a consistent triplet, and $H$ is the B\"acklund transformation between $F$ and $G$. Note that a large class of consistent triplets was given in  \cite{ZZ-JNMP-2018,ZZ-FMC-2019} where $F$ and/or $G$ are not necessarily  affine-linear. 
Such type of triplets also played an important role in generating rational solutions for certain lattice equations ({\em cf}. \cite{ddZZ}).

Similarly, by using the transformation \eqref{eq:psii}, the relation \eqref{3.10} can be easily reduced to
\begin{equation}
 V\wt V+U \wt U=-i\delta/s,\quad  V\h V+U \h U=-i\delta/t,
\end{equation}
which form the B\"acklund transformations for two copies of the H3($\delta$) equations for $V$ and $U$.

\subsection{Ground state solutions}

 It is interesting enough to look at the ``ground state solutions'', corresponding to the {\em zero-soliton} of the lpKdV equation \eqref{eq:H1}
\begin{equation}\label{eq:zerosol}
  u = -\alpha n-\beta m +c,
\end{equation}
where $c$ is a constant.
According to \eqref{eq:1}, one has  $h=-2\alpha$, $g= \alpha-\beta$, which yields
\begin{equation}
  \vp_1 =(\alpha+p)^n(\beta+p)^m\,,\quad   \vp_2 =(\alpha-p)^n(\beta-p)^m\,,
\end{equation}
as the ground states for the Lax pair \eqref{eq:dschreq}.   

Consider, for instance,  the lpmKdV($\delta$) equation \eqref{eq:lpmkdvd1}. Assuming that $\alpha,\beta$ and $p$ are real, we can parameterize $\alpha, \beta$ and $p$ as
 \begin{equation}
   \label{eq:32}
 \alpha= s \cosh( \gamma'),\quad  p = s \sinh ( \gamma')\,,\quad \beta= t \cosh(\lambda'),\quad p = t \sinh(\lambda')\,.
 \end{equation}
By taking account of the branches \eqref{sign1}, one has
 \begin{equation}
   \label{eq:7}
 v_{1} =  s^{-n}t^{-m}\vp_{1} =\gamma^n \lambda^m \,, \quad
 v_{2} = s^{-n}t^{-m}\vp_{2}=(-1)^{n+m}\gamma^{-n} \lambda^{-m}\,,
\end{equation}
where $\gamma = e^{\gamma'}, \lambda = e^{\lambda'} $.
It is clear  now the general expressions of the  ground state  solutions to  the lpmKdV($\delta$) equation is  in the form
 \begin{equation}
 v = (A+(-1)^{n+m}C)v_1+ (B+(-1)^{n+m}D)v_2\,,
\end{equation}
where $A, B, C, D$ are constants, and $\delta = 4p (AD-BC) $.

Let  $C=D=0$,  then $\delta=0$,  and $v$ as a linear combination of  $v_1$, $v_2$ solves the  lpmKdV equation.  Without loss of generality, we also assume that both $ \gamma, \lambda$ are greater than $1$, then    $v_{1} \to \infty$, as $n,m\to \infty$, and $v_{1} \to 0$, as $n,m\to -\infty$, while  $v_{2}$ represents an ``oscillatory'' term.

Let   $B=C=0$, then  $\delta = 4p AD$ is nonzero, and one has
\begin{equation}
\label{v}
v = A \gamma^n \lambda^m+ D\gamma^{-n} \lambda^{-m}\,,
\end{equation}
  which satisfies the lpmKdV($\delta$) equation with $\delta = 4p AD$. Now, $v$ is behaves asymptotically different from $v_1$, since $v\to \infty$ as $n,m\to \pm\infty$.
The term  $\delta$ in  \eqref{v} can be regarded as a product of interactions between $v_1$ and $ (-1)^{n+m}v_{2} $. An inverse-scattering transform for initial-value problems of lpmKdV($\delta$) subject to different asymptotic behaviors is required to make the analysis rigorous.

\section{Lattice eigenfunction KdV equations: case II }

We continue to develop lattice eigenfunction KdV equations by taking account of another Lax pair\begin{equation}
  \label{eq:dschreqab}
  \widetilde{\widetilde{\phi}} +h\widetilde{\phi}+\alpha^2\,\phi = q^2
  \phi\,,       \quad      \widehat{\phi} =\widetilde{\phi}-g\phi \,,
\end{equation}
where $h,g$ are the same functions as appeared in \eqref{eq:dschreq}.  We assume, in general, that $p^2\neq q^2$. Then, we obtain the following two quad-equations as eigenfunction equations
\begin{equation}
  \label{eq:lee2}
s^2\mu^2y \widehat{y} +\widetilde{y} \widehat{\widetilde{y}} - t^2 \nu^2y \widetilde{y} -\widehat{y} \widehat{\widetilde{y}} - (\alpha^2-\beta^2)(y \widehat{\widetilde{y}} +  \widetilde{y}\widehat{y} )=0\,,
\end{equation}
and
\begin{equation}
  \label{eq:lee2d}
s^2\mu^2Y \widehat{Y} +\widetilde{Y} \widehat{\widetilde{Y}} - t^2 \nu^2Y \widetilde{Y} -\widehat{Y} \widehat{\widetilde{Y}}=(\alpha^2-\beta^2)\left(Y \widehat{\widetilde{Y}} +  \widetilde{Y}\widehat{Y} +\delta (p^2-q^2)^2 s^{2n}\mu^{2n} t^{2m} \nu^{2m}\right)\,,
\end{equation}
where $\delta$ is a constant, and
\begin{equation}
\label{eq:sstt}  s^2=\alpha^2 - p^2\,,\quad \mu^2= \alpha^2 - q^2\,,  \quad t^2=\beta^2 - p^2\,,\quad \nu^2= \beta^2 - q^2\,.
\end{equation}
We call \eqref{eq:lee2} lattice eigenfunction KdV-II equation (leKdV-II) and  \eqref{eq:lee2d} leKdV-II($\delta$) equation. LeKdV-II($\delta$)  contains leKdV-II as a special case with $\delta =0$. They are $3$D-consistent and $\ZZ_2$-symmetric, and in connection to  the Q3 and Q3($\delta$)  equations.
With suitable reductions, the LeKdV-II($\delta$) equation can be reduced to the Q1($\delta$) equation.
\subsection{Derivations}
\begin{proposition}
\label{prop:41} Let $\varphi$ and $\phi$ be solutions of the Lax pairs \eqref{eq:dschreq} and \eqref{eq:dschreqab} respectively, with $q^2\neq p^2$, then, a function $y$ defined by
    \begin{equation}\label{eq:y1}
  y = \varphi \widetilde{\phi} - \phi\widetilde{\vp}
\end{equation}
solves the leKdV-II equation \eqref{eq:lee2}.
  \end{proposition}
  \prf
It follows from the forms of the Lax pairs \eqref{eq:dschreq} and \eqref{eq:dschreqab} that
  \begin{equation} \label{eq:y2}
 \widetilde{\vp}  \widetilde{\widetilde{\phi}}   -  \widetilde{\phi}\widetilde{\widetilde{\vp}}    = s^2 \vp\widetilde{\phi}-\mu^2 \phi\widetilde{\vp}\,,\end{equation}
  where $s, \mu$  are defined in \eqref{eq:sstt}.  The left-hand side is a shift of $y$ in the ~$\widetilde{~}$~ direction, which implies
  \begin{equation}
    \widetilde{y} =  s^2 \vp\widetilde{\phi}-\mu^2 \phi\widetilde{\vp}\,.
  \end{equation}
The above expressions allow to express the quantities $ \vp\widetilde{\phi}$ and $\phi\widetilde{\vp}$ in terms of $y$ and $\widetilde{y}$:
 \begin{equation}
   \label{eq:y3}
    \vp\widetilde{\phi} = \frac{\widetilde{y}-\mu^2 y }{s^2-\mu^2}\,,\quad     \phi\widetilde{\vp} = \frac{\widetilde{y}-s^2 y }{s^2-\mu^2}\,,
  \end{equation}

  One also has
  \begin{equation}
y=      \varphi \widehat{\phi} -  \phi\widehat{\vp}\,, \quad     \widehat{y} =  t^2 \vp\widehat{\phi}-\nu^2 \phi\widehat{\vp}\,,
  \end{equation}
which allows to express the quantities  $ \vp\widehat{\phi} $  and $ \phi\widehat{\vp}$ in terms of $y$ and $\widehat{y}$ similar to \eqref{eq:y3} by replacing $(~\widetilde{~}~,s,\mu) $  by  $(~\widehat{~}~,t,\nu) $. Using the consistency relation
  \begin{equation}\label{eq:cvp1}
      (\vp\widetilde{\phi})\, \widehat{(\phi \widetilde{\vp} )} = ( \vp\widehat{\phi}) \widetilde{({\phi}\widehat{\vp})}\,,
    \end{equation}
one obtains  an equation for $y$ which is the leKdV-II equation \eqref{eq:lee2}.
  \finprf
\medskip

The consistency relations \eqref{eq:cvp1} are consequences of Prop.~\ref{prop:1}.  Let $\vp_1, \vp_2$ (resp. $\phi_1, \phi_2$) be a  solution basis of \eqref{eq:dschreq} (resp. \eqref{eq:dschreqab}), 
then one has the general form of $y$  using $\vp = C_1\vp_1+C_2\vp_2$, 
$\phi = D_1\phi_1+D_2\phi_2$, where $C_j, D_j$, $j=1,2$, are constants,
\begin{equation}
  \label{ycd}
  y = C_1 D_1 \,y_{11}+C_1 D_2\, y_{12}+C_2 D_1\, y_{21}+C_2 D_2\, y_{22}\,,\quad y_{ij} = \varphi_i \widetilde{\phi}_j - \phi_j\widetilde{\vp}_i\,.
\end{equation}
Such $y$ is a special linear combination of $y_{ij}$, while each $y_{ij}$ is a solution to \eqref{eq:lee2} as well. A generic linear combination of  $y_{ij}$ will satisfy the leKdV-II($\delta$) equation.
  \begin{proposition}
 Let $\varphi_1, \varphi_2$ (resp. $\phi_1, \phi_2$) be two linearly independent solutions of the Lax pair \eqref{eq:dschreq} (resp. \eqref{eq:dschreqab}), with $q^2\neq p^2$. Let $Y$ be defined as
 \begin{equation}
   \label{eq:dY}
  Y = \bra{\Phi_p} M \ket {\widetilde{\Phi}_q}-\bra{\widetilde{\Phi}_p} M\ket {{\Phi}_q}\,,
\end{equation}
where
\begin{equation}
  \bra{\Phi_p} = \bma \vp_1 &  \vp_2 \ema \,, \quad   \ket{\Phi_q} = \bma \phi_1 \\  \phi_2 \ema \,, \quad M =  \bma A &  B \\ C & D \ema\,,
\end{equation}
$A,B,C,D$ are constants,  then $Y$ satisfies the leKdV-II$(\delta)$ equation, where
\begin{equation}
  \delta =\rho_1 \rho_2 \det{M}\,,
\end{equation}
where $\rho_1, \rho_2$ are constants of integration from
\begin{equation}
 |\vp_1,\vp_2|_{\widetilde{~}} =\rho_1\, s^{2n}t^{2m}\,,\quad   |\phi_1,\phi_2|_{\widetilde{~}}= \rho_2\,  \mu^{2n}\nu^{2m}\,,
\end{equation}
  \end{proposition}
  \prf
The proof is similar  to that of Prop.~\ref{prop:41}. Here, $\bra{\cdot}$ denotes a row vector which is the transpose of $\ket{\cdot}$, \ie $\bra{\cdot} = \ket{\cdot}^\intercal$. One has $\bra{\Phi_p}$ and $\ket{\Phi_q}$ satisfying
  \begin{equation}
    \bra{\widetilde{\widetilde{\Phi}}_p}+h\bra{\widetilde{\Phi}_p}+s^2 \bra{\Phi_p} =0\,,\quad  \ket{\widetilde{\widetilde{\Phi}}_q}+h\ket{\widetilde{\Phi}_q}+\mu^2 \ket{\Phi_q} = 0\,.
  \end{equation}
  Right multiplying the first equation  by $M \ket{\widetilde{\Phi}_q}$, 
  and left multiplying the second equation by  $\bra{\widetilde{\Phi}_p}M $, 
  and making the difference, one gets
  \begin{equation}
\label{yy}\bra{\widetilde{\Phi}_p} M \ket {\widetilde{\widetilde{\Phi}}_q}  -\bra{\widetilde{\widetilde{\Phi}}_p} M \ket {\widetilde{\Phi}_q} = s^2 \bra{\Phi_p}M\ket {\widetilde{\Phi}_q} -\mu^2 \bra{\widetilde{\Phi}_p}M\ket {\Phi_q}\,.
\end{equation}
Due to the definition of $Y$, this implies
\begin{equation}
  \widetilde{Y} =  s^2 \bra{\Phi_p}M\ket {\widetilde{\Phi}_q} -\mu^2 \bra{\widetilde{\Phi}_q}M\ket {\Phi_p}\,,
\end{equation}
which allows to express the quantities $\bra{\Phi_p}M\ket {\widetilde{\Phi}_q}$ and $\bra{\widetilde{\Phi}_p}M\ket {\Phi_q}$ in terms of $Y$ and $\widetilde{Y}$. Moreover,  $\bra{\Phi_p}$ and $\ket{\Phi_q}$ also satisfy
\begin{equation}
\bra{\widehat{\Phi}_p} =   \bra{\widetilde{\Phi}_p} -g   \bra{{\Phi}_p}\,,  \quad  \ket{\widehat{\Phi}_q} =   \ket{\widetilde{\Phi}_q} -g   \ket{{\Phi}_q}\,,
\end{equation}
meaning
\begin{equation}
  Y =\bra{\Phi_p} M \ket {\widehat{\Phi}_q}-\bra{\widehat{\Phi}_p} M\ket {{\Phi}_q}\,,
\end{equation}
and that the quantities $\bra{\Phi_p} M \ket {\widehat{\Phi}_q}$ and $\bra{\widehat{\Phi}_p} M\ket {{\Phi}_q}$ can be expressed in terms of $Y$ and $\widehat{Y}$. Then, we have  the following consistency relations
\begin{align}\label{fucom}
\frac{ \text{E}(Y)}{(p^2-q^2)^2}= & \bra{\Phi_p}M\ket {\widetilde{\Phi}_q}  \left(\widehat{\bra{{\widetilde{\Phi}}_p}M\ket {{\Phi}_q}}\right)^\intercal-  \bra{\Phi_p}M\ket {\widehat{\Phi}_q}  \left(\widetilde{\bra{\widehat{\Phi}_p}M\ket {\Phi_q}} \right)^\intercal\nonumber \\
= & \bra{\Phi_p}M \left(\ket {\widetilde{\Phi}_q}  \bra{\widehat{\Phi}_q} -\ket {\widehat{\Phi}_q}  \bra{\widetilde{\Phi}_q}   \right) M^\intercal\ket {\widehat{\widetilde{\Phi}}_p} \nonumber \\
  =& (\det{M})\, G\,F\,,
\end{align}
where  E($Y$)$=0$ is the leKdV-II equation  for $Y$ and
\begin{equation}
G=\widetilde{\phi}_1\widehat{\phi}_2-\widetilde{\phi}_2\widehat{\phi}_1\,,\quad F= \vp_1 \widehat{\widetilde{\vp}}_2-\vp_2 \widehat{\widetilde{\vp}}_1\,.
\end{equation}
Again following Prop.~\ref{prop:1} and Lemma~\ref{lem1}, one has
\begin{equation}
  G  = g  |\phi_1,\phi_2|_{\widetilde{~}}=g \, \rho_2 \mu^{2n}\nu^{2m}\,,\quad F= -(h+\widetilde{g})  |\vp_1,\vp_2|_{\widetilde{~}} =-(h+\widetilde{g})\, \rho_1 s^{2n}t^{2m}\,,
\end{equation}
where $\rho_1, \rho_2$ are constants of integration. Knowing the lattice KdV equation \eqref{eq:10sys}, one completes the proof.
\finprf
\medskip

Note that the adjoint Lax pairs involving $\vps$ and $\phi^\ast$ can also be used to constructed eigenfunction equations. These yield very similar equations as leKdV-II($\delta$).

Expanding \eqref{eq:dY} yields
\begin{equation}
\label{gY}  Y = A \,y_{11}+B\, y_{12}+C\, y_{21}+ D\, y_{22}\,, \quad y_{ij} = \varphi_i \widetilde{\phi}_j - \phi_j\widetilde{\vp}_i\,,
\end{equation}
which, in contrast to \eqref{ycd}, is a true linear combination of $y_{ij}$, and can be seen as the general form of exact solutions of leKdV-II($\delta$), where $\{y_{ij}\}$ form the solution basis. The term $\delta$ is a result of interactions between $y_{11}$ and $y_{22}$, and between $y_{12}$ and $y_{21}$.
\subsection{Connections to Q3($\delta$) and Q1($\delta$) equations}
Now we connect the leKdV-II($\delta$) equation to some well-known $3$D-consistent quad-equations. As usual, the branches of the parameters $s,t$ (resp. $\mu,\nu$) are fixed with the sign of $p$ (resp. $q$)
\begin{equation}
  \text{sign} \,s(p) = \text{sign}\,  p\,, \quad   \text{sign} \,t(p) = \text{sign}\,  p\,,
  \quad \text{sign} \,\mu(q) = \text{sign}\,  q\,, \quad   \text{sign} \,\nu(q) = \text{sign}\,  q\,.
\end{equation}
\begin{itemize}
\item Connection to NQC($\delta$) equation: let $Y$ satisfy the  leKdV-II($\delta$) equation \eqref{eq:lee2d}, and
\begin{equation}
  Y  = (p-q)s_+^{n}\mu_+^nt_+^{m}\nu_+^m \left((p+q)S -1\right)\,, \quad S:=S(n,m;\alpha,p,\beta,q)\,,
\end{equation}
where
\begin{equation}
  s_\pm =\alpha \pm p\,,\quad   \mu_\pm =\alpha \pm q\,,  \quad t_\pm =\beta \pm p\,, \quad  \nu_\pm =\beta \pm q\,,
\end{equation}
then $S$ satisfies a $\delta$-extension of the NQC equation
\begin{equation}
  \label{eq:lpmkdvd}
 \text{NQC}(S)=\frac{\delta(\alpha^2-\beta^2)}{s_+\mu_+t_+\nu_+}\frac{s^n_-\mu_-^nt_-^m\nu_-^m}{s^n_+\mu_+^nt_+^m\nu_+^m}\,,
\end{equation}
where
\begin{equation}
\text{NQC}(S)=( 1 - \mu_+ \widetilde{S} + s_- S) (1 - s_+\widehat{\widetilde{S}} + \mu_- \widehat{S}) - (1 -\nu_+\widehat{S} + t_- S)(1 - t_+  \widehat{\widetilde{S}} + \nu_- \widetilde{S})\,,
\end{equation}
and  NQC($S$)$=0$ is the NQC equation \cite{NQC} for $S$ without $\delta$.
\item Connection to  Q3($\delta$): let  $Y$ satisfy the  leKdV-II($\delta$) equation \eqref{eq:lee2d}, and
  \begin{equation}
  Y  =  s^{n}\mu^nt^{m}\nu^m X\,, \quad X:=X(n,m;\alpha,p,\beta,q)\,,
\end{equation}
then $X$ satisfies
\begin{equation}\label{eq:Q3dpara}
  s\,\mu(X\widehat{X}+\widetilde{X}\widehat{\widetilde{X}})
  -t\,\nu(X\widetilde{X}+\h{X}\widehat{\widetilde{X}}) =  (\alpha^2-\beta^2)\left(\widetilde{X}\widehat{X}+X\widehat{\widetilde{X}}+(p^2-q^2)^2\frac{\delta}{s\mu t\nu} \right)\,,
\end{equation}
which is the Q3($\delta$) equation \cite{ABS1, ABS2}\footnote{Using the parametrization $A = \frac{\mu}{s}$, $B= \frac{\nu}{t}$, it follows from \eqref{eq:Q3dpara} that
\begin{equation}
  \left(
   A-\frac{1}{A} \right)(X\widetilde{X}+\widetilde{X}\widehat{\widetilde{X}}) -\left(
   B-\frac{1}{B} \right)(X\widehat{X}+\h{X}\widehat{\widetilde{X}}) = \left(
   \frac{A}{B}-\frac{B}{A} \right) \left(\widetilde{X}\widehat{X}+X\widehat{\widetilde{X}}+\delta(
   A-\frac{1}{A} )(
   B-\frac{1}{B} ) \right)\,,
\end{equation}
which is the original form of Q3($\delta$) presented in \cite{ABS1}.}
\end{itemize}
\begin{rmk}
Exact solutions of $Q3(\delta)$ have been obtained in \cite{SQ3, NAH}
 for soliton solutions, and in \cite {NA-elliptic} for elliptic soliton solutions, following the Cauchy matrix approach. In those derivations,  explicit analytic structures (discrete exponential-type functions for soliton solutions, and elliptic functions for elliptic solutions) are presumed. However, our constructions of  the Q3($\delta$) equation, and the related NQC($\delta$) and leKdV-II($\delta$) equation, do not require  knowledge of the analytic natures of the solution basis, \ie $y_{ij}$, $i,j=1,2$, as shown in \eqref{gY}. Moreover, we give an adequate explanation of the term $\delta$, which can be interpreted as interactions between different solution basis $y_{ij}$.  \end{rmk}
We conclude this section, by providing the connection of leKdV-II($\delta$) to the Q1($\delta$)  equation.
\begin{proposition}
 Let $\varphi_1, \varphi_2$ be two linearly independent solutions of the Lax pair \eqref{eq:dschreq}. Let $\chi$ be defined as
 \begin{equation}
   \label{eq:dY}
  \chi = A \,x_{11}+B( x_{12}+ x_{21})+ D\, x_{22}\,,
\end{equation}
where $A,B,D$ are constants, and
\begin{equation}
x_{ij} = \vp_i \widetilde{\vp}_{j,p}-\widetilde{\vp}_j \vp_{i,p}\,,\quad i,j = 1,2\,,
\end{equation}
with  $\vp_{\ell,p}$, $\ell=1,2$, denoting partial derivative of $ \vp_\ell $ with respect to $p$,  then $\chi$ satisfies
\begin{equation}\label{eq:EQ}
  s^4 \chi \widehat{ \chi} +\widetilde{\chi} \widehat{\widetilde{ \chi}}-t^4\chi \widetilde{ \chi}- \widehat{\chi}\widehat{\widetilde{  \chi}}  = (\alpha^2-\beta^2)\left(\chi\widehat{\widetilde{  \chi}}+\widetilde{\chi}\widehat{ \chi}+ 4p^2\delta s^{4n}t^{4m}\right)\,,
\end{equation}
where $s,t$ are defined in \eqref{eq:sstt} and
\begin{equation}
   \delta =\rho^2( A D -B^2)\,,
\end{equation}
with $\rho$ being the constant of integration as appeared in \eqref{eq:9}.
\end{proposition}
  \prf
  Consider the leKdV-II($\delta$) equation \eqref{eq:lee2d} and its solutions \eqref{gY}. Let $C=B$ and $q =  p+\epsilon$. Taking the limit $\epsilon \to 0$, then $Y$ in \eqref{gY} is reduced to
  \begin{equation}
    \lim_{\epsilon\to 0}Y\vert_{  p =q+\epsilon} = \epsilon\,\chi +o(\epsilon)\,.
  \end{equation}
It can be easily checked that $\chi$ satisfies the equation \eqref{eq:EQ} by apply the limit  $q =  p+\epsilon$,  $\epsilon \to 0$ to  leKdV-II($\delta$).  \finprf
  \medskip

Clearly, the equation \eqref{eq:dY} is a degenerate case of leKdV-II($\delta$). It is connected to Q1($\delta$) through
the following transformation \begin{equation}
 \chi  =  s^{2n}t^{2m}Z\,, \quad Z:=Z(n,m;s,t)\,,
\end{equation}
where $s,t $ are defined in \eqref{eq:sstt}, and $Z$ satisfies the  Q1($\delta$) equation \cite{ABS1, ABS2}
\begin{equation}
  s^2(Z -  \widetilde{Z}) (\widehat{Z}-\widehat{\widetilde{Z}})
  -t^2(Z - \h{Z})(\widetilde{Z} - \widehat{\widetilde{Z}}) =  4p^2\delta\left(\frac{1}{t^2}-\frac{1}{s^2}\right).
\end{equation}
The exact solution structures are thus
\begin{equation}
  Z = s^{-2n}t^{-2m}(A \,x_{11}+B( x_{12}+ x_{21})+ D\, x_{22} )\,,
\end{equation}
where $x_{ij}$ is defined in \eqref{ycd}, and $\delta \propto AD -B^2$.


\section{Concluding remarks}\label{sec-5}

In this paper, based on the Lax formulations of the lattice KdV equation, we develop the KdV-type lattice eigenfunction equations, which enables us to make a precise connection to the ABS lattice equations with a $\delta$ term, namely the H3($\delta$), Q1($\delta$) and Q3($\delta$) equations.
The appearance of the $\delta$  term is the result of interactions between different eigenfunctions.
By construction, we give exact solution structures of these equations in terms of eigenfunctions.
This also helps to understand the ABS classification and the associated lattice KdV-type equations. The discrete Schr\"odinger equation and its discrete deformation  \eqref{eq:dschreq} are the fundamental objects in determining the equations and their solutions: the lpKdV (H1) and H2 can be considered as potential lattice KdV equations, since they can be expressed in terms of potential variable $u$ and its Miura transformation, {\em cf}. \cite{ZhPeZh},  other equations (excepting Q4)  are related to the eigenfunction equations leKdV($\delta$) \eqref{H3eqdd} and leKdV-II($\delta$) \eqref{eq:lee2} (considering Q2 as a degeneration of Q3($\delta$)).

Note that  a variety of exact solutions to the ABS lattice equations, including rational \cite{ddZZ, ZZ}, soliton \cite{NAH, HZ}, elliptic soliton \cite{NA-elliptic} and algebraic-geometric \cite{CZ}  solutions, has been found,  using different approaches. The solution structures of the ABS lattice equations with a $\delta$ term presented in this paper, in particular, the solutions of Q3($\delta$), 
allow us to provide precise descriptions of exact solutions to all equations, 
excepting Q4, from the ABS list. One natural continuation of the present work is to establish a rigorous spectral analysis of the Lax pair \eqref{eq:dschreq} subject to various boundary/asymptotic conditions, and provide a unified picture of solution structures for the ABS equations in the inverse scattering transform scheme. Note that the inverse scattering transform for certain ABS equations was investigated in \cite{BJ} in view of soliton solutions.

Our approach to constructing integrable lattice equations represents an alternative to some existing methods such as the consistency approach \cite{ABS1, ABS2}, or the Cauchy matrix approach \cite{NAH}. Its effectiveness lies in the fact that some apparently complicated nonlinear lattice equations can be fully understood and operated at the level of linear problems. It also naturally accommodates the equations with a $\delta$ term.  We believed that our approach can be well-adapted to the task of searching/extending lattice integrable models in the higher-rank analogs of the lattice KdV equation along the discrete Gel'fand-Dikii hierarchy \cite{GD}. This will be shown in future work.

%

\appendix

\section{Darboux-Crum formulae for the Lax pair  \eqref{eq:dschreq}}\label{app:1}
Here we present the main formulae concerning exact solutions of the  Lax pair  \eqref{eq:dschreq} generated by Darboux transformations. Proofs can be found in \cite{ZhPeZh}. 
Two types of exact solutions, related to soliton and elliptic soliton solutions of the lpKdV equation, are also provided. 
The most general algebraic-geometric solutions can be found, for instance,  in \cite{CZ}.
These formulae can be directly used to generate  soliton and elliptic solutions of the lattice equations presented in this paper.

\subsection{Darboux-Crum formulae}

A one-step Darboux transformation for the Lax pair  \eqref{eq:dschreq} can be expressed as
\begin{equation}
  \varphi \mapsto \varphi[1]=\widetilde{\varphi}-\gamma \varphi\,,\quad   h \mapsto h[1] =\widetilde{h} -\widetilde{\widetilde{\gamma}}-\gamma\,,
\end{equation}
where $\gamma = \widetilde{\varphi}_1\varphi^{-1}_1$ with $\varphi_1$ being a particular solution of  \eqref{eq:dschreq} with  $p^2=p_1^2$.

The Crum's formulae are compact expressions of the action of an $N$-step Darboux transformation. Assuming there exist $N$ linearly independent solutions $\varphi_j$  of  the Lax pair  \eqref{eq:dschreq}  associated with $p^2 = p^2_j$, $j = 1\,,2\,,\dots\,,N$.  Then  the $N$-step Darboux transformation amounts to the  map
  \begin{equation}
    \varphi\mapsto  \varphi[N] =  \frac{|\varphi_1,\varphi_2,\dots, \varphi_N, \varphi|_{\widetilde{~}}  }{|\varphi_1,\varphi_2,\dots, \varphi_N|_{\widetilde{~}} } \,,
   \quad
h\mapsto  h[N] =  h^{(N)} -r^{(2)}+r \,.
\end{equation}
The superscript $^{(N)}$ denotes $N$  shifts in the ~$\widetilde{}$~ direction, and  $|\varphi_1,\varphi_2,\dots, \varphi_N|_{\widetilde{~}} $ is the Casorati determinant for functions $\varphi_1\,, \varphi_2\,,\, \dots\,,\, \varphi_N$, namely,
\begin{equation}
  \label{CASO}
 |\varphi_1,\varphi_2,\dots, \varphi_N|_{\widetilde{~}}  = \begin{vmatrix} \varphi_1 & \varphi_1^{(1)} & \dots & \varphi_1^{(N-1)} \\
\vdots &\vdots & \dots & \vdots   \\
\varphi_N & \varphi_N^{(1)} & \dots & \varphi_N^{(N-1)} \end{vmatrix} \, ,
\end{equation}
and $r$ is in the form
\begin{equation}
  \label{eq:HN1}
  r= - \frac{\begin{vmatrix} \vp_1 & \vp_1^{(1)} & \dots & \vp_1^{(N-2)} & \vp_1^{(N)}  \\
\vdots &\vdots & \dots & \vdots& \vdots   \\
\vp_N & \vp_N^{(1)} & \dots & \vp_N^{(N-2)} & \vp_N^{(N)} \end{vmatrix}}{|\varphi_1,\varphi_2,\dots, \varphi_N|_{\widetilde{~}} }\,.
\end{equation}

\subsection{Solutions of the Lax pair  \eqref{eq:dschreq}}\label{app:1-2}
Soliton solutions can be constructed using the zero-soliton solutions of lpKdV
\[u=-\alpha n-\beta m +c,\]
\ie \eqref{eq:zerosol}. Then, one has
\begin{equation}
\varphi= \rho^{+}(\alpha+p)^n(\beta+p)^m + \rho^{-}(\alpha-p)^n(\beta-p)^m,
\end{equation}
as general solutions to the Lax pair  \eqref{eq:dschreq},  where $\rho^{\pm}$ are arbitrary constants. The particular solutions $\varphi_j$, $j=1,2, \cdots, N$,  are in the form $\varphi_j=\varphi|_{p=p_j,\rho^{\pm}=\rho_j^{\pm}}$.

The lpKdV equation \eqref{eq:H1} also allows elliptic solutions, {\em cf}.~\cite{NA-elliptic}. Consider the Weierstrass elliptic functions $\sigma(z), \zeta(z), \wp(z)$ with the same
fundamental period parallelogram.
When $\alpha, \beta, p, q$ are parameterized as
\begin{equation}
\alpha^2=\wp(a),\quad \beta^2=\wp(b),\quad
p^2=\wp(\kappa),\quad q^2=\wp(\varepsilon),
\end{equation}
lpKdV has a solution
\begin{equation}
  u = \zeta(\xi) -\zeta(a) n-\zeta(b) m + c_0,
\end{equation}
where $\xi = n a +m b+\xi_0$ and $c_0, \xi_0$ are constants.
The Lax pair  \eqref{eq:dschreq} turns out to be
\begin{subequations}
  \label{eq:dschreq-ell}
\begin{align}
  \widetilde{\widetilde{\varphi}} +
(\zeta(\xi+2a)-\zeta(\xi)-2\zeta(a)) \wt{\varphi}+(\wp(a)-\wp(\kappa) )\varphi =&0, \label{eq:dschreq-ella}   \\
\h{\varphi} - \wt{\varphi}
+(\zeta(\xi+b)-\zeta(\xi+a) -\zeta(b) +\zeta(a) )\varphi =&0 \label{eq:dschreq-ellb},
\end{align}
\end{subequations}
in which \eqref{eq:dschreq-ella} can be considered as a discrete analog of the  Lam\'e equation.
Solutions to \eqref{eq:dschreq-ell}  are
\begin{equation}\label{phi-ell}
\varphi= \rho^{+}\Phi_{-\kappa}^\xi (\Phi_\kappa^a)^n(\Phi_\kappa^b)^m
+ \rho^{-}\Phi_{\kappa}^\xi (\Phi_{-\kappa}^a)^n(\Phi_{-\kappa}^b)^m,
\end{equation}
where
\begin{equation}
\Phi_x^y=\frac{\sigma(x+y)}{\sigma(x)\sigma(y)}\,.
\end{equation}
Note that the following elliptic identities are needed to derive the above formulae
\begin{align}
 \zeta(x)+\zeta(y)+\zeta(z)-\zeta(x+y+z)=
\frac{\sigma(x+y)\sigma(y+z)\sigma(x+z)}{\sigma(x)\sigma(y)\sigma(z)\sigma(x+y+z)},\quad  \wp(x)-\wp(y)=\Phi_{x}^{y}\Phi_{x}^{-y}.
\end{align}

%
%

\end{document}